# Une nouvelle étape pour l'informatique : Binaire, Biologique, Quantique

Xavier Vasques

IBM Technology, France

## Abstract

Le centre de données de demain est un centre de données composé de systèmes hétérogènes, qui tourneront des charges de travail hétérogènes. Les systèmes hétérogènes seront dotés d'accélérateurs binaires, d'inspiration biologique et quantique.  Ces architectures seront les fondations pour relever les défis non résolus aujourd'hui avec des applications dans tous les secteurs allant de la recherche à l'industrie en passant par des applications ayant un réel impact sociétal. Tel un chef d'orchestre, le cloud hybride permettra grâce à une couche de sécurité et d'automatisation intelligente de mettre ces systèmes en musique. Cet article fait un tour d'horizon des technologies actuelles et futures permettant de surpasser les limites de la loi de Moore.

# Introduction

L'informatique classique a connu des progrès remarquables guidés par la loi de Moore. Celle-ci nous dit que tous les deux ans, nous doublons le nombre de transistors dans un processeur et nous augmentons par la même occasion les performances par deux ou réduisons les coûts par deux. Cette cadence a ralenti au cours de la dernière décennie et nous assistons aujourd'hui à un plafonnement. Ce ralentissement oblige à une transition. Nous devons repenser l'informatique et notamment aller vers des architectures de systèmes hétérogènes dotés d'accélérateurs afin de répondre au besoin de performances dans des enveloppes de coûts traditionnelle (1). Les progrès qui ont été faits sur la puissance de calcul brute nous ont néanmoins amenés à un point où les modèles de calcul d'inspiration biologique basés sur des réseaux de neurones sont désormais hautement considérés dans l'état de l'art (2) (3) (4). L'intelligence artificielle (IA) est également un domaine amenant des opportunités de progrès, mais aussi des défis. Les capacités de l'IA ont considérablement augmenté dans leur capacité à interpréter et analyser des données. L'IA peut également être exigeante en termes de puissance de calcul à cause de la complexité de certains « workflows ». L'IA peut également être appliqué à la gestion et à l'optimisation de systèmes informatiques entiers (1). En parallèle des accélérateurs classiques ou d'inspiration biologique, l'informatique quantique programmable émerge grâce à plusieurs dizaines d'années d'investissement dans la recherche dans un but de surmonter les limites physiques traditionnelles. Cette nouvelle ère de l'informatique va potentiellement avoir les capacités de faire des calculs aujourd'hui non possibles par des ordinateurs classiques mais aussi non possibles pour les ordinateurs classiques, binaires, de demain. Les futurs systèmes devront intégrer des capacités d'informatiques quantiques pour effectuer des calculs spécifiques. La recherche avance rapidement. IBM a mis à disposition des ordinateurs quantiques programmables pour la première fois sur le cloud en mai 2016 et a annoncé son ambition de doubler le volume quantique chaque année. On parle de la loi de Gambetta. Le cloud est aussi un élément qui apporte des défis et des opportunités considérables dans l'informatique. Le cloud est capable de fournir des capacités de calcul complexe et géographiquement répartie dans l'économie mondiale connectée. Le cloud a un rôle important à jouer. Les centres de données de demain seront pilotés par le cloud et équipés de systèmes hétérogènes qui tourneront des « workloads » hétérogènes de manière sécurisé. Les données ne seront plus centralisées ou décentralisées mais seront organisées en « hub ». La puissance de calcul devra se trouver à la source de la donnée qui aujourd'hui arrive à des volumes extrêmes. Les systèmes de stockage sont aussi pleins de défis pour favoriser la disponibilité, la performance, la gestion mais aussi la fidélité des données. Nous devons concevoir des architectures permettant d'extraire de la valeur des données de plus en plus complexes et souvent régulées qui posent de multiples défis, notamment la sécurité, le chiffrement, la confidentialité ou la traçabilité. L'avenir de l'informatique sera construit avec des systèmes hétérogènes composés d'informatique classique dite systèmes binaires ou bits, d'informatique d'inspiration biologique et d'informatique quantique dite systèmes quantiques ou qubits (1). Ces composants hétérogènes seront orchestrés et déployés par une structure cloud hybride qui masque la complexité tout en permettant l'utilisation et le partage sécurisés des systèmes et des données privés et publics.

## 1. Les systèmes binaires

Les premiers ordinateurs binaires furent construits dans les années 40 : Colossus (1943) puis ENIAC (IBM - 1945). Colossus a été conçu pour déchiffrer des messages secrets allemands et l'ENIAC conçu pour calculer des trajectoires balistiques. L'ENIAC (acronyme de l'expression anglaise Electronic Numerical Integrator And Computer) est en 1945 le premier ordinateur entièrement électronique construit pour être « Turing-complet » : il peut être reprogrammé pour résoudre, en principe, tous les problèmes calculatoires. L'ENIAC a été programmé par des femmes, dites les « femmes ENIAC ». Les plus célèbres d'entre elles étaient Kay McNulty, Betty Jennings, Betty Holberton, Marlyn Wescoff, Frances Bilas et Ruth Teitelbaum. Ces femmes avaient auparavant effectué des calculs balistiques sur des ordinateurs de bureau mécaniques pour l'armée. L'ENIAC pèse alors 30 tonnes, occupe une surface de 72 m2 et consomme 140 kilowatts. Quelle que soit la tâche effectuée par un ordinateur, le processus sous-jacent est toujours le même : une instance de la tâche est décrite par un algorithme qui est traduit en une suite de 0 et de 1, pour donner lieu à l'exécution dans le processeur, la mémoire et les dispositifs d'entrée/sortie de l'ordinateur. C'est la base du calcul binaire qui en pratique repose sur des circuits électriques dotés de transistors pouvant être dans deux modes : « ON » permettant au courant de passer et « OFF » le courant ne passe pas. À partir de ces 0 et ces 1 on a donc développé au cours des 80 dernières années une théorie de l'information classique construite à partir d'opérateurs booléens (XAND, XOR), de mots (Octets), et une arithmétique simple basée sur les opérations suivantes : « 0+0=0, 0+1=1+0=1, 1+1=0 (avec une retenue), et vérifier si 1=1 , 0=0 et 1≠0. Bien entendu à partir de ces opérations, il est possible de construire des opérations beaucoup plus complexes que les ordinateurs peuvent effectuer des millions de milliards de fois par seconde pour les plus puissants d'entre eux. Tout cela est devenu tellement « naturel » que l'on oublie totalement que chaque transaction sur un serveur informatique, sur un PC, une calculette, un smartphone se décompose en ces opérations binaires élémentaires. Dans un ordinateur, ces 0 et 1 sont contenus dans des « BInary digiTs » ou « bits » qui représentent la plus petite quantité d'information contenue dans un système informatique. L'ingénieur en génie électrique et mathématicien Claude Shannon (1916-2001) fut l'un des pères fondateurs de la théorie de l'information. Pendant 20 ans, Claude Shannon a travaillé au Massachusetts Institute of Technology (MIT) et parallèlement à ses activités académiques, il a travaillé aux laboratoires Bell. Il se marie en 1949 à Madame Moore. Pendant la Seconde Guerre mondiale, Claude Shannon travaille pour les services secrets américains, en cryptographie, afin de localiser des messages cachés dans des codes allemands. Une de ses contributions essentielles concerne la théorie de la transmission des signaux (5) (6) (7). C'est dans ce contexte qu'il mit au point une théorie de l'information notamment en comprenant que toute donnée, même la voix ou des images, peut se transmettre à l'aide d'une suite de 0 et de 1.

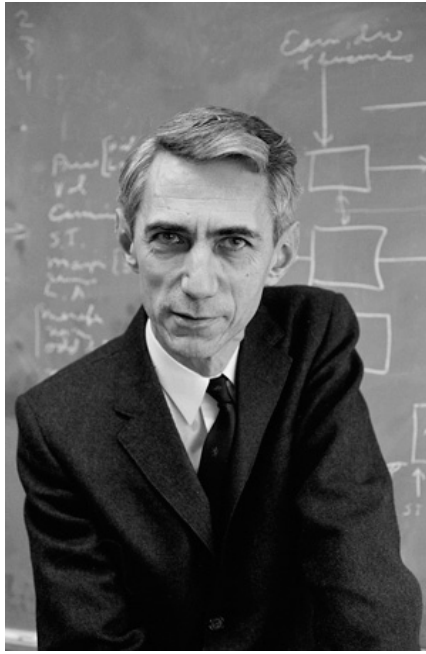
Photo: Alfred Eisenstaedt/The LIFE Picture Collection/Getty Images (40)

Le binaire utilisés par les ordinateurs classiques est apparu au milieu du 20e siècle, lorsque les mathématiques et l'information ont été combinées d'une manière nouvelle pour former la théorie de l'information, lançant à la fois l'industrie informatique et les télécommunications. La force du binaire réside dans sa simplicité et sa fiabilité. Un bit est soit zéro soit un, un état qui peut être facilement mesuré, calculé, communiqué ou stocké (1). Lorsqu'ils sont fournis avec la même entrée binaire, les programmes classiques produisent toujours la même sortie. Cela nous a permis de créer des systèmes incroyablement robustes et fiables pour gérer des charges de travail à volume élevé. Grâce à cette méthode, différents systèmes de calculs et de stockage de données ont vu le jour jusqu'au stockage de données numériques sur une molécule d'ADN (8).

Nous avons aujourd'hui des exemples de systèmes binaires avec des possibilités incroyables. Un premier exemple, c'est l'IBM Z. Un processeur IBM Z, Single Chip Module (SCM), utilise une technologie Silicon On Insulator (SOI) à 14 nm. Il contient 9.1 milliards de transistors. Il y a 12 cœurs par PU SCM à 5.2GHz. Cette technologie permet avec un seul et même système de pouvoir traiter 19 milliards de transactions encryptées par jour et 1000 milliards de transactions web par jour. L'ensemble des IBM Z installés dans le monde traitent aujourd'hui 87% des transactions par carte bancaire et huit mille milliards de payements par an (9).

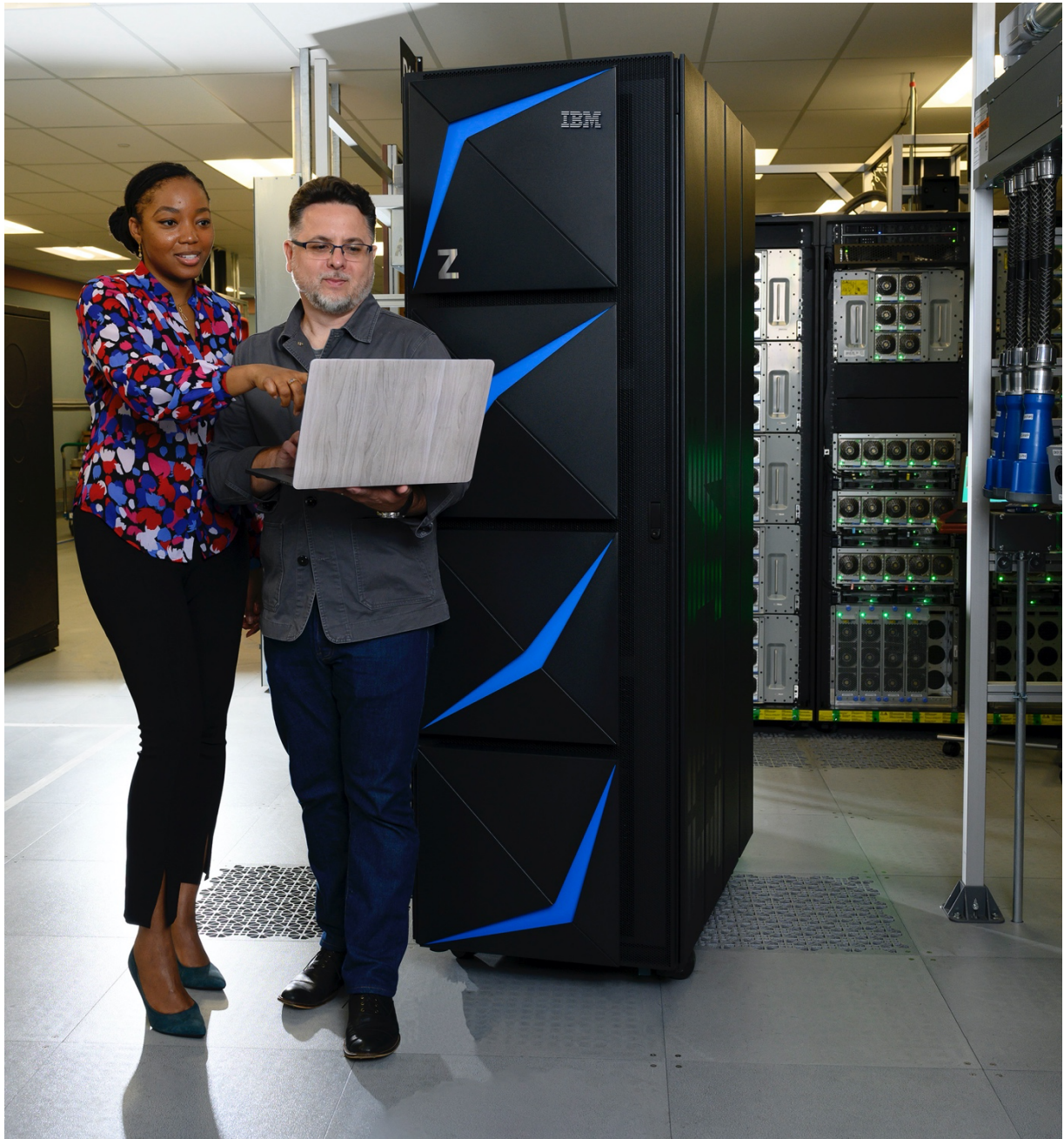
Source : IBM News Room (41)

Nous pouvons aussi citer deux ordinateurs parmi les plus puissants du monde, respectivement nommés Summit et Sierra. Ils sont installés dans le Laboratoire d'Oak Ridge dans le Tennessee et dans le Laboratoire National Lawrence à Livermore en Californie. Ces ordinateurs aident à modéliser des supernovas ou de nouveaux matériaux, des chercheurs les utilisent pour trouver des solutions contre le cancer, étudier la génétique et l'environnement. Summit est capable de délivrer une puissance de 200 pétaflops avec une capacité de stockage de 250 petabytes. Il est composé de 9216 IBM Power9 CPUs, 27648 NVIDIA Tesla GPUs et une communication réseau de 25 Gigabytes par seconde entre les nœuds. Malgré tout, même l'ordinateur le plus puissant du monde, doté d'accélérateurs GPUs, ne peut pas tout calculer.

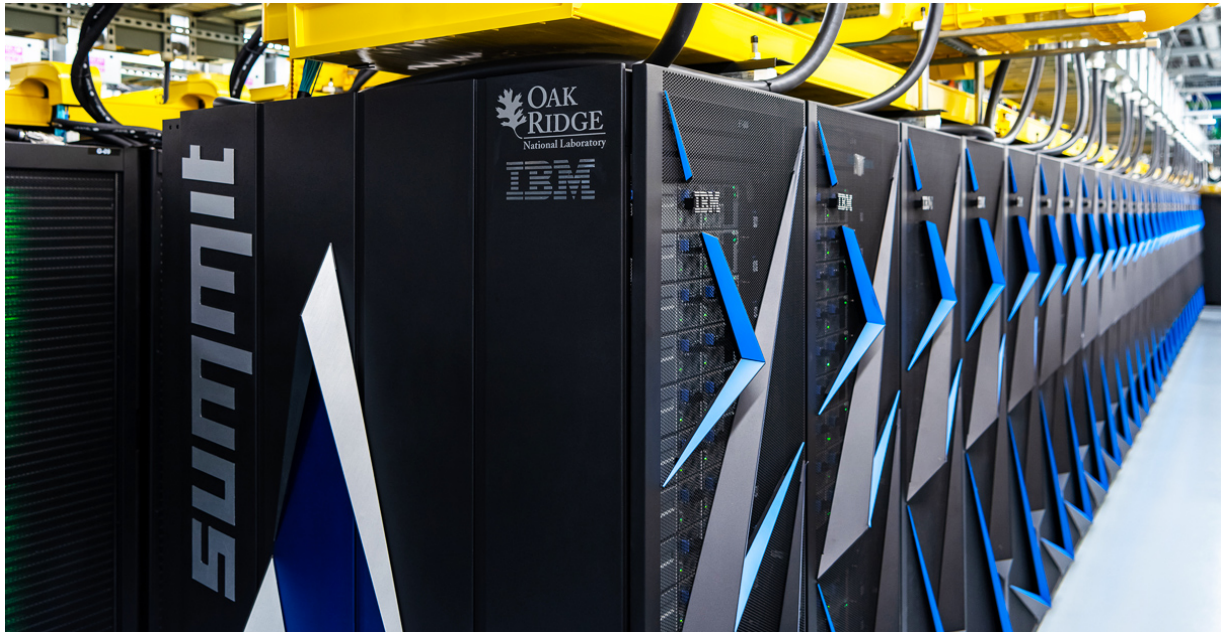
Source : IBM News Room (42)

Aujourd'hui, ce type de technologie est indispensable à la recherche médicale. Et nous l'avons vu pendant cette crise du Covid-19. Nous pouvons prendre l'exemple de l'utilisation de la puissance des supercalculateurs avec le consortium HPC COVID-19 (https://covid19-hpc-consortium.org). C'est un effort public-privé initié par plusieurs institution dont IBM visant à mettre la puissance de supercalculateurs à la disposition des chercheurs travaillant sur des projets liés à la COVID-19 pour les aider à identifier des thérapies potentielles à court terme pour les patients atteints par le virus.

Depuis son lancement en mars 2020, la capacité de calcul du Consortium a presque doublé pour atteindre 600 pétaflops (million de milliards d'opérations flottantes par secondes), contre 330 pétaflops en mars.

Ensemble, le Consortium a contribué à soutenir de nombreux projets de recherche, notamment comprendre combien de temps les gouttelettes respiratoires persistent dans l'air. Cette recherche d'une équipe de l'Université d'État de l'Utah a simulé la dynamique des aérosols à l'intérieur, offrant un aperçu de la durée pendant laquelle les gouttelettes respiratoires persistent dans l'air. Ils ont constaté que les gouttelettes provenant de la respiration persistent dans l'air beaucoup plus longtemps qu'on ne le pensait auparavant, en raison de leur petite taille par rapport aux gouttelettes de la toux et des éternuements. Un autre projet concerne la recherche sur la réutilisation de médicaments pour des traitements potentiels. Un projet d'une équipe de la Michigan State University a examiné les données d'environ 1600 médicaments approuvés par la FDA pour voir s'il existe des combinaisons possibles qui pourraient aider à traiter le COVID-19. Ils ont trouvé prometteurs au moins deux médicaments approuvés par la FDA: la proflavine, un désinfectant contre de nombreuses bactéries, et la chloroxine, un autre médicament antibactérien.

En France, une collaboration entre l'Institut Pasteur et IBM France est un autre exemple montrant le besoin d'accélération (43). Comme nous pouvons avoir des milliers de molécules candidates pour un traitement thérapeutique potentiel, l'utilisation de systèmes accélérés et du deep learning permet de filtrer les meilleures correspondances afin de proposer une sélection de composés chimiques capables de s'attacher aux protéines des agents pathogènes. En faisant cela, le processus de conception des médicaments pourrait être considérablement accéléré. L'IA aidera également les chercheurs à mieux profiler les interactions protéine-protéine impliquées dans le développement des pathologies, ainsi qu'à mieux comprendre la dynamique des infections dans les cellules humaines. Grâce à cette approche innovante, le cycle de développement d'un traitement thérapeutique pourrait être accéléré, passant potentiellement de plusieurs années à quelques mois, voire quelques semaines, tout en économisant des millions d'euros.

Les ordinateurs, les smartphones et leurs applications, internet que l'on utilise dans nos vies de tous les jours fonctionnent avec des 0 et des 1. Le binaire couplé à loi de Moore, un héritage de 50 ans, a permis de construire des systèmes robustes et fiables. Pendant 50 ans, nous avons vu des évolutions incrémentales, linéaires, pour avoir des gains de performance. Les prochaines années vont nous amener leurs lots d'innovations afin d'avoir des gains de performance notamment au niveau des matériaux, des processus de contrôle ou méthode de gravure : on parle de transistor tridimensionnel, lithographie extrême ultraviolet ou de nouveau matériaux comme le hafnium ou germanium. Le binaire continue donc à évoluer et va jouer un rôle central dans le centre de donnée de demain.

Récemment, IBM a fait un grand pas en avant dans la technologie des puces en fabriquant la première puce à 2 nm permettant de presser 50 milliards de transistors sur une puce de la taille d'un ongle. L'architecture peut aider les fabricants de processeurs à améliorer les performances de 45 % avec la même quantité d'énergie que les puces 7 nm actuelles, ou le même niveau de performances en utilisant 75 % d'énergie en moins. Les appareils mobiles dotés de processeurs basés sur 2 nm pourraient avoir jusqu'à quatre fois la durée de vie de la batterie de ceux dotés de chipsets 7 nm. Les ordinateurs portables bénéficieraient d'une augmentation de la vitesse de ces processeurs, tandis que les véhicules autonomes détecteront et réagiront plus rapidement aux objets. Cette technologie profitera à l'efficacité énergétique des centres de données, à l'exploration spatiale, à l'intelligence artificielle, à la 5G et 6G et à l'informatique quantique.

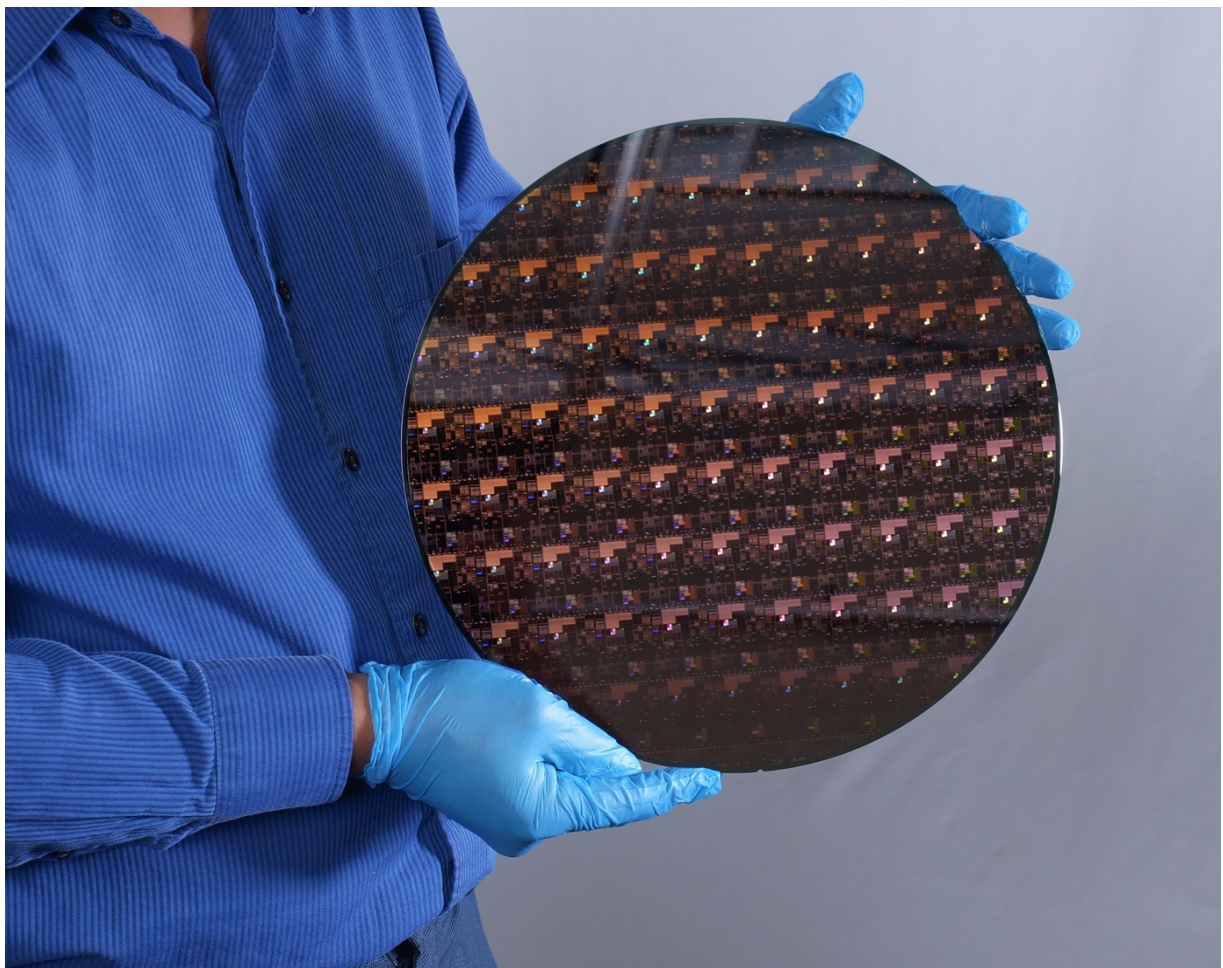

Source : IBM News Room (https://newsroom.ibm.com/2021-05-06-IBM-Unveils-Worlds-First-2-Nanometer-Chip-Technology,-Opening-a-New-Frontier-for-Semiconductors)

Malgré les commentaires sur les limites de la loi de Moore et les goulots d'étranglement des architectures actuelles, les innovations continuent et le binaire jouera un rôle central dans le centre de donnée de demain. Malgré tout, certains défis ne pourront pas être adressés uniquement avec le binaire. Pour ces défis, nous avons besoin de repenser l'informatique avec de nouvelles approches en nous inspirant de la nature comme la biologie et la physique.

## 2. Les systèmes d'inspiration biologique

Une molécule d'ADN est destinée par nature à stocker de l'information génétique grâce aux quatre bases azotées qui composent une molécule d'ADN (A, C, T, et G). Il est aujourd'hui possible de transcrire des données numériques en un nouveau code génétique. Le séquençage ADN permet ensuite de lire l'information stockée. L'encodage est lui automatisé via un logiciel. Une molécule d'ADN est composée de 3 milliards de nucléotides (base azotée). Une équipe à récemment publié dans le journal Science la capacité de stocker un système d'exploitation, un film français de 1895 (L'Arrivée d'un *train* à La *Ciotat par Louis Lumière)*, un article scientifique, une photo, un virus et une carte cadeau de 50$ dans des brins d'ADN, et de récupérer les données sans erreurs. Dans un gramme d'ADN, 215 pétaoctets de données peuvent être stockés. Il serait possible de stocker toutes les données créées par les humains dans une seule pièce. De plus, l'ADN peut théoriquement conserver des données en parfait état pendant une durée extrêmement longue. Dans des conditions idéales, on estime que l'ADN pourrait encore être déchiffré après plusieurs millions d'années grâce aux « gènes de longévité ». L'ADN peut résister aux conditions météorologiques les plus extrêmes. Ce n'est pas encore pour demain, car il existe encore de nombreux défis comme par exemple les coûts élevés et les délais de traitement qui peuvent être extrêmement longs.

Inspirés par les neurosciences et combinant la biologie et l'information, les systèmes de deep learning ont été conçus pour atteindre et même dépasser les performances humaines sur de nombreuses tâches. Ces systèmes peuvent par exemple utiliser des réseaux neuronaux pour créer des modèles. Ces modèles sont formés en apprenant à partir de grands ensembles de données. L'expression IA en tant que telle apparaît en 1956. Plusieurs chercheurs américains, dont John McCarthy, Marvin Minsky, Claude Shannon et Nathan Rochester d'IBM, très en pointe dans des recherches qui utilisent des ordinateurs pour autre chose que des calculs scientifiques, se sont réunis à l'université de Dartmouth, aux États-Unis. Trois ans après le séminaire de Dartmouth, les deux pères de l'IA, McCarthy et Minsky, fondent le laboratoire d'IA au MIT. Il y a eu beaucoup d'investissement, une trop grande ambition, imiter le cerveau humain, et beaucoup d'espoir non concrétisé à l'époque. Les promesses n'ont pas été tenues. Une approche plus pragmatique est apparue dans les années 70 et 80 avec l'émergence du machine learning et la réapparition des réseaux de neurones à la fin des années 80. Cette approche plus pragmatique, l'augmentation de la puissance de calcul et de la quantité de données a permis qu'aujourd'hui l'IA s'invite dans tous les domaines, c'est un sujet transversal. L'utilisation massive d'IA pose quelques défis comme le besoin de labéliser les données à notre disposition. Le problème avec l'automatisation c'est qu'elle demande beaucoup de travail manuel. L'IA a besoin d'éducation. Cela est fait par des dizaines de milliers de travailleurs autour du monde ce qui ne ressemble pas vraiment à ce que l'on pourrait appeler une vision futuriste. Un autre défi c'est le besoin en puissance de calcul. L'IA a besoin d'être entrainée et pour cela l'IA est de plus en plus gourmande en moyens de calculs. L'entrainement demande un doublement des capacités de calcul tous les 3.5 mois (10).

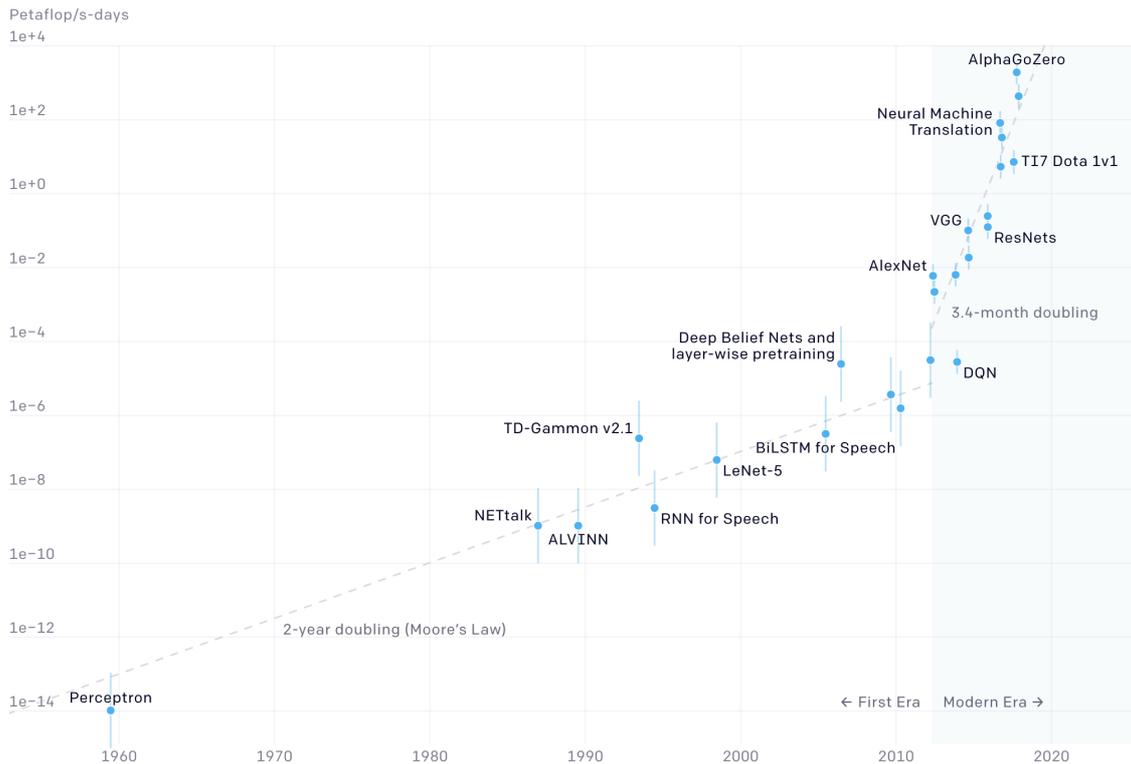

**Source:** AI and Compute, OpenAI, https://openai.com/blog/ai-and-compute/#fn1

Plusieurs approches sont aujourd'hui utilisées et envisagées. Comme pour le supercalculateur Summit, le calcul de certaines charges de travail est déporté vers des accélérateurs comme les GPUs. Il en existe d'autres comme les FPGAs (Field Programmable Gate Arrays ou "réseaux logiques programmables) qui peuvent réaliser la ou les fonctions numériques voulues. L'intérêt est qu'une même puce peut être utilisée dans de nombreux systèmes électroniques différents.

Le progrès dans le domaine des neurosciences va permettre de désigner des processeurs directement inspirés du cerveau. La façon que notre cerveau transmet l'information n'est pas binaire. Et c'est grâce à Santiago Ramón y Cajal (1852-1934), histologiste et neuroscientifique espagnol, prix Nobel de physiologie ou médecine en 1906 avec Camillo Golgi que nous connaissons mieux l'architecture du système nerveux. Les neurones sont des entités cellulaires séparées par de fins espaces, les synapses, et non des fibres d'un réseau ininterrompu (11). L'axone d'un neurone transmet des influx nerveux, des potentiels d'actions, à des cellules cibles. La prochaine étape en ce qui concerne le développement de nouveaux types de processeurs spécialisés pour l'IA et inspirés par le cerveau c'est de penser différemment la manière dont on calcul aujourd'hui. Aujourd'hui un des problèmes majeurs de la performance c'est le mouvement des données entre les différents composants de l'architecture de von Neumann : processeur, mémoire, stockage. Il est donc impératif d'ajouter des accélérateurs analogiques. Ce qui domine aujourd'hui les calculs numériques et notamment les calculs de deep learning est la multiplication en virgule flottante. Une des méthodes envisagées comme un moyen efficace de gagner en puissance est de revenir en arrière en réduisant la précision également appelée calcul approximatif. Par exemple, les moteurs de précision 16 bits sont plus de 4x plus petits que les moteurs de précision 32 bits. Ce gain augmente les performances et l'efficacité énergétique. En termes simples, dans le calcul approximatif, nous pouvons faire un compromis en échangeant la précision numérique contre l'efficacité du calcul. Certaines conditions sont néanmoins nécessaires comme développer en parallèle des améliorations algorithmiques pour garantir une isoprécision (1). Dans le cadre d'algorithmes ML/DL, IBM a récemment démontré le succès de cette approche avec des nombres en virgules flottantes à 8 bits, en utilisant de nouvelles techniques pour maintenir la précision des calculs de gradients et la mise à jour des poids pendant la rétropropagation (12) (13). De même pour l'inférence d'un modèle issu de l'entrainement d'algorithme de deep learning, l'utilisation unique de l'arithmétique entière sur 4 ou 2 bits de précision atteint une précision comparable à une gamme de modèles de réseaux de neurones et ensembles de données populaires (14). Cette progression conduira à une augmentation spectaculaire de la capacité de calcul pour les algorithmes de deep learning au cours de la prochaine décennie.

Les accélérateurs analogiques sont une autre voie évitant le goulot d'étranglement de l'architecture de von Neumann (15) (16). L'approche analogique utilise des unités de traitement résistives programmables (RPUs) non volatiles qui peuvent encoder les poids d'un réseau neuronal. Des calculs comme la multiplication matricielle ou vectorielle ou les opérations des éléments matricielles peuvent être effectués en parallèle et en temps constant, sans mouvement des poids (1). Dans une architecture de puce analogique, le réseau IA est alors représenté par des tableaux liés entre eux. Des fonctions d'activation non linéaires sont insérées dans la connexion entre les tableaux et peuvent être effectuées soit dans l'espace numérique, soit en analogique. Les poids étant stationnaires, le trafic de données est considérablement réduit, ce qui atténue le goulot d'étranglement de von Neumann. Cependant, contrairement aux solutions numériques, l'IA analogique sera plus sensible aux propriétés des matériaux et intrinsèquement plus sensible au bruit et à la variabilité. Ces facteurs doivent être traités par des solutions architecturales, des nouveaux circuits et algorithmes. Par exemple, les mémoires non volatiles (NVM) analogues (17) peuvent efficacement accélérer les algorithmes de « backpropagation ». En combinant le stockage à long terme dans des dispositifs de mémoire à changement de phase (PCM), la mise à jour quasi linéaire des condensateurs CMOS conventionnels et des nouvelles techniques pour éliminer la variabilité d'un appareil à l'autre, des résultats significatifs ont commencé à émerger pour le calcul de DNNs (Deep Neural Network) (18) (19) (20). Ces expériences ont utilisé une approche mixte matérielle logicielle, combinant des simulations logicielles d'éléments de système faciles à modéliser avec précision (tels que des appareils CMOS) avec une implémentation matérielle complète des composants PCM. La recherche s'est également lancée dans une quête afin de construire une puce directement inspirée du cerveau (21). Dans un article publié dans Science (22), IBM et ses partenaires universitaires ont mis au point un processeur appelé SyNAPSE qui est composé d'un million de neurones. La puce ne consomme que 70 milliwatts et est capable d'effectuer 46 milliards d'opérations synaptiques par seconde, par watt, littéralement un superordinateur synaptique tenant dans une paume de main. Nous sommes passés des neurosciences aux superordinateurs, à une nouvelle architecture informatique, à un nouveau langage de programmation, à des algorithmes, à des applications et maintenant à une nouvelle puce qui s'appelle TrueNorth (23). TrueNorth c'est un circuit intégré CMOS neuromorphique produit par IBM en 2014. Il s'agit d'un réseau de processeur « manycore », avec 4096 cœurs, chacun ayant 256 neurones simulés programmables pour un total d'un peu plus d'un million de neurones. À son tour, chaque neurone possède 256 synapses programmables permettant le transport des signaux. Par conséquent, le nombre total de synapses programmables est légèrement supérieur à 268 millions. Le nombre de transistors de base est de 5,4 milliards. Étant donné que la mémoire, le calcul et la communication sont gérés dans chacun des 4096 cœurs neurosynaptiques, TrueNorth contourne le goulot d'étranglement de l'architecture von Neumann et est très économe en énergie. Il a une densité de puissance de 1/10 000 des microprocesseurs conventionnels.

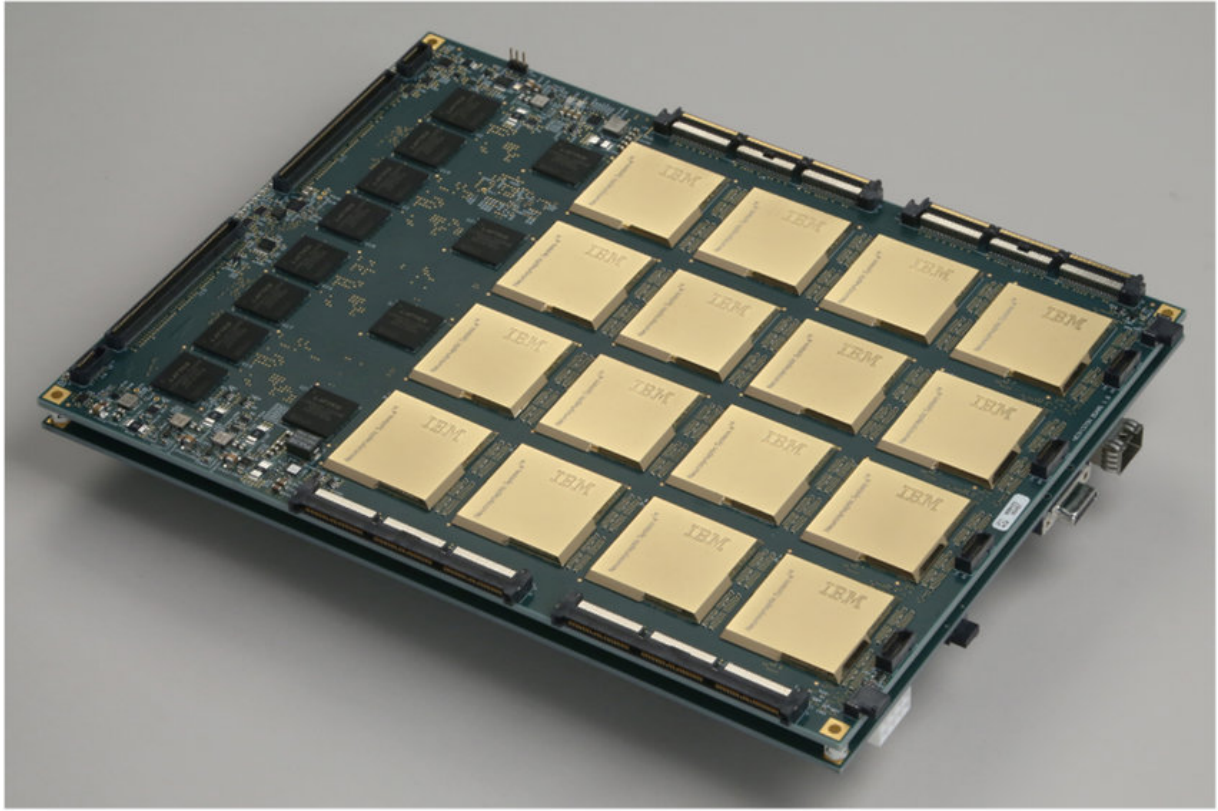

Source : https://www.research.ibm.com/articles/brain-chip.shtml

### 3. Les systèmes quantiques : Qubits

Les lois de la physique nous aident aussi à imaginer les ordinateurs de demain. Par exemple, dans un article publié dans Nature, les physiciens et ingénieurs d'IBM ont décrit comment ils ont réussi l'exploit d'écrire et de lire des données dans un atome d'holmium. C'est un pas en avant symbolique mais qui prouve que cette approche fonctionne et que nous pourrions un jour avoir un stockage de données atomique. Pour faire un parallèle pour montre ce que cela signifie, imaginez que nous puissions stocker toute la bibliothèque iTunes de 35 millions de chansons sur un appareil de la taille d'une carte de crédit. Dans l'article, les nanoscientifiques ont démontré la capacité de lire et d'écrire un bit de données sur un atome. À titre de comparaison, les disques durs d'aujourd'hui utilisent 100 000 à un million d'atomes pour stocker un seul bit d'information. Bien sûr, nous ne pouvons pas éviter la discussion autour de l'informatique quantique. Les bits quantiques - ou qubits - combinent la physique avec l'information et sont les unités de base d'un ordinateur quantique. Les ordinateurs quantiques utilisent des qubits dans un modèle de calcul basé sur les lois de la physique quantique. Des algorithmes quantiques convenablement conçus sont capables de résoudre des problèmes de grande complexité en exploitant la superposition quantique et l'intrication pour accéder à un espace d'état exponentiel, puis en amplifiant la probabilité de calculer la bonne réponse par une interférence constructive. C'est à partir du début des années 1980, sous l'impulsion du physicien et prix Nobel Richard Feynman que germe l'idée de la conception et du développement d'ordinateurs quantiques : Là où un ordinateur « classique » fonctionne avec des bits de valeurs 0 ou 1, l'ordinateur quantique utilise les propriétés fondamentales de la physique quantique et repose sur des « quantum bits (qubits) ». Au-delà de ce progrès technologique, l'informatique quantique ouvre la voie au traitement de tâches informatiques dont la complexité est hors de portée de nos ordinateurs actuels. Mais reprenons depuis le début.

Au début du XXe siècle, les théories de la physique dite « classique » sont dans l'impossibilité d'expliquer certains problèmes observés par les physiciens. Elles doivent donc être reformulées et enrichies. Sous l'impulsion des scientifiques, elle va évoluer dans un premier temps vers une « nouvelle mécanique » qui deviendra « mécanique ondulatoire » et finalement « mécanique quantique ». La mécanique quantique est la théorie mathématique et physique qui décrit la structure fondamentale de la matière et l'évolution dans le temps et dans l'espace des phénomènes de l'infiniment petit. Une notion essentielle de la mécanique quantique est la dualité « onde corpuscule ». Jusqu'aux années 1890, les physiciens considèrent que le monde est composé par deux types d'objets ou de particules :  d'une part celles qui ont une masse (comme les électrons, les protons, les neutrons, les atomes ...), et d'autre part celles qui n'en ont pas (comme les photons, les ondes ...). Pour les physiciens de l'époque, ces particules sont régies par les lois de la mécanique newtonienne pour celles qui ont une masse et par les lois de l'Électromagnétisme pour les ondes. Nous disposions donc de deux théories de la « Physique » pour décrire deux types d'objets différents. La mécanique quantique invalide cette dichotomie et introduit l'idée fondamentale de la dualité onde corpuscule. Les particules de matière ou les ondes doivent être traitées par les mêmes lois de la physique. C'est l'avènement de la mécanique ondulatoire qui deviendra quelques années plus tard la mécanique quantique. De grands noms sont associés au développement de la mécanique quantique comme Niels Bohr, Paul Dirac, Albert Einstein, Werner Heisenberg, Max Planck, Erwin Schrödinger et bien d'autres. Max Planck et Albert Einstein, en s'intéressant au rayonnement émis par un corps chauffé et à l'effet photoélectrique, furent les premiers à comprendre que les échanges d'énergie lumineuse ne pouvaient se faire que par « paquet ». D'ailleurs, Albert Einstein obtient le prix Nobel de physique suite à la publication de sa théorie sur l'aspect quantifié des échanges d'énergie en 1921. Niels Bohr étendit les postulats quantiques de Planck et d'Einstein de la lumière à la matière, en proposant un modèle reproduisant le spectre de l'atome d'hydrogène. Il obtient le prix Nobel de physique en 1922, en définissant un modèle de l'atome qui dicte le comportement des quanta de lumière. En passant d'un palier d'énergie à un autre inférieur, l'électron échange un quantum d'énergie. Pas à pas, des règles furent trouvées pour calculer les propriétés des atomes, des molécules et de leurs interactions avec la lumière.

De 1925 à 1927, toute une série de travaux de plusieurs physiciens et mathématiciens donna corps à deux théories générales applicables à ces problèmes :
- La mécanique ondulatoire de Louis de Broglie et surtout de Erwin Schrödinger ;
- La mécanique matricielle de Werner Heisenberg, Max Born et Pascual Jordan.

Ces deux mécaniques furent unifiées par Erwin Schrödinger du point de vue physique, et par John von Neumann du point de vue mathématique. Enfin, Paul Dirac formula la synthèse ou plutôt la généralisation complète de ces deux mécaniques, que l'on nomme aujourd'hui la mécanique quantique. L'équation fondamentale de la mécanique quantique est l'équation de Schrödinger.

L'informatique quantique a débuté lors d'une petite conférence, désormais célèbre, en 1981, organisée conjointement par IBM et le MIT, sur la physique de l'informatique. Le physicien lauréat du prix Nobel Richard Feynman a mis au défi les informaticiens d'inventer un nouveau type d'ordinateur basé sur des principes quantiques, afin de mieux simuler et prédire le comportement de la matière réelle (24): "Je ne suis pas satisfait de toutes les analyses qui vont avec juste la théorie classique, parce que la nature n'est pas classique, bon sang, et si vous voulez faire une simulation de la nature, vous feriez mieux de la faire avec la mécanique quantique..."

La matière, a expliqué Feynman, est constituée de particules telles que des électrons et des protons qui obéissent aux mêmes lois quantiques qui régiraient le fonctionnement de ce nouvel ordinateur. Depuis, les scientifiques se sont attaqués au double défi de Feynman : comprendre les capacités d'un ordinateur quantique et trouver comment en construire un. Les ordinateurs quantiques seront très différents des ordinateurs d'aujourd'hui, non seulement en ce à quoi ils ressemblent et en quoi ils sont faits, mais, plus important encore, dans ce qu'ils peuvent faire. Nous pouvons également citer une phrase célèbre de Rolf Landauer, un physicien qui a travaillé chez IBM : « L'information est physique ». Les ordinateurs sont bien sûr des machines physiques. Il faut donc tenir compte des coûts énergétiques engendrés par des calculs, l'enregistrement et la lecture des bits d'information ainsi que des dissipations d'énergie sous forme de chaleur. Dans un contexte où les liens entre thermodynamique et information étaient, l'objet de bien des interrogations, Rolf Landauer, a cherché à déterminer la quantité minimale d'énergie nécessaire pour manipuler un unique bit d'information dans un système physique donné. Il y aurait donc une limite, aujourd'hui appelée limite de Landauer et découvert en 1961, qui définit que tout système informatique est obligé de dissiper un minimum de chaleur et donc consommer un minimum d'électricité. Ces recherches sont fondamentales, car elles montrent que tout système informatique dispose d'un seuil thermique et électrique minimum que l'on ne pourra pas dépasser. Cela signifiera que l'on arrivera à la consommation minimum d'une puce informatique et qu'elle ne pourra pas dégager moins d'énergie. Ce n'est pas pour tout de suite, mais les scientifiques expliquent que cette limite sera surtout présente lors de la conception de puces quantiques. Les travaux récents de Charles Henry Bennett chez IBM ont consisté en un réexamen des bases physiques de l'information et l'application de la physique quantique aux problèmes des flux d'informations. Ses travaux ont joué un rôle majeur dans le développement d'une interconnexion entre la physique et l'information.

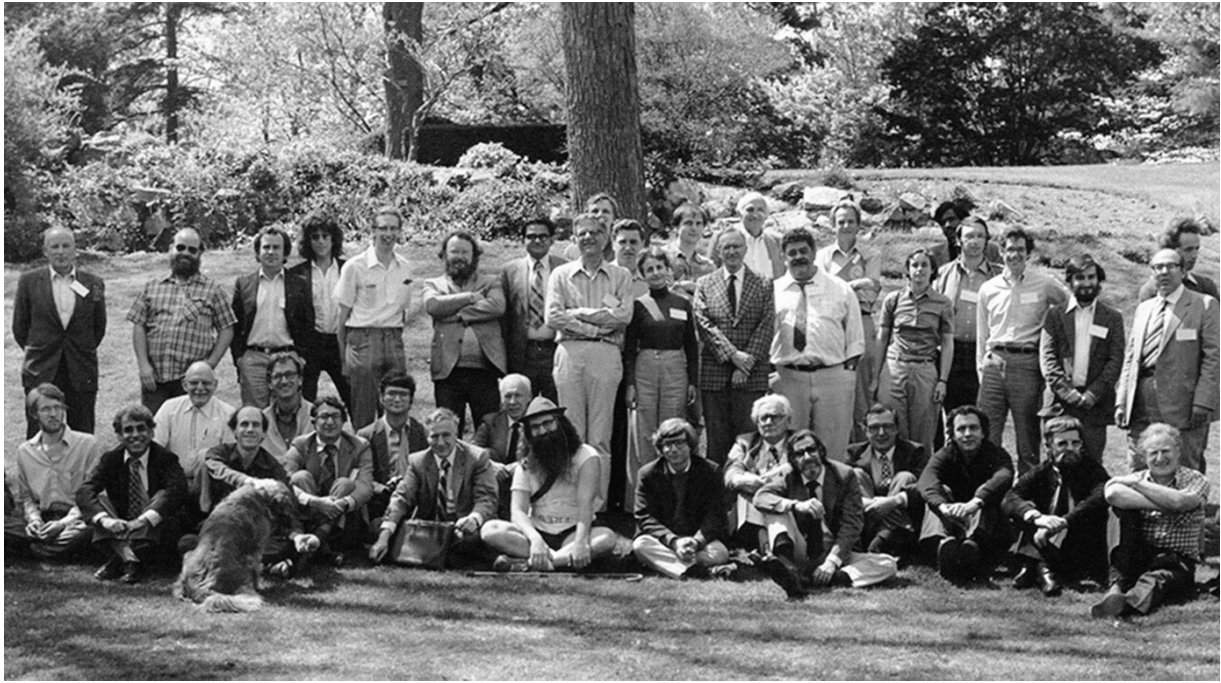
Source : https://research.ibm.com/blog/qc40-physics-computation

Pour un ordinateur quantique le qubit est l'entité de base, représentant, à l'instar du « bit » la plus petite entité permettant de manipuler de l'information. Il possède deux propriétés fondamentales de la mécanique quantique : superposition & intrication. Un objet quantique (à l'échelle microscopique) peut exister dans une infinité d'états (tant qu'on ne mesure pas cet état). Un qubit peut donc exister dans n'importe quel état entre 0 et 1. Les qubits peuvent prendre à la fois la valeur 0 et la valeur 1, ou plutôt « une certaine quantité de 0 et une certaine quantité de 1 », comme une combinaison linéaire de deux états notés |0> et |1>, avec les coefficients **α** et **β**. Donc là où un bit classique ne décrit « que » 2 états (0 ou 1), le qubit peut en représenter une « infinité ». C'est un des avantages potentiels du calcul quantique du point de vue de la théorie de l'information. On peut se faire une idée de la superposition d'états en utilisant l'analogie du ticket de loterie : un ticket de loterie est soit gagnant, soit perdant une fois que l'on connaît le résultat du jeu. Par contre, avant le tirage, ce ticket n'était ni gagnant ni perdant. Il avait simplement une certaine probabilité d'être gagnant et une certaine probabilité d'être perdant, il était en quelque sorte gagnant et perdant à la fois. Dans le monde quantique, toutes les caractéristiques des particules peuvent être sujettes à cette indétermination : par exemple, la position d'une particule est incertaine. Avant la mesure, la particule n'est ni au point A, ni au point B. Elle a une certaine probabilité d'être au point A et une certaine probabilité d'être au point B. Cependant, après la mesure, l'état de la particule est bien défini : elle est au point A ou au point B.

L'intrication est une autre propriété étonnante de la physique quantique. Lorsque l'on considère un système composé de plusieurs qubits, il peut leur arriver de « lier leur destin » c'est-à-dire de ne pas être indépendants l'un de l'autre même s'ils sont séparés dans l'espace (alors que les bits « classiques » sont complètement indépendants les uns des autres). C'est ce que l'on appelle l'intrication quantique. Si l'on considère un système de deux qubits intriqués alors la mesure de l'état d'un de ces deux qubits nous donne une indication immédiate sur le résultat d'une observation sur l'autre qubit.

Pour illustrer naïvement cette propriété on peut là aussi utiliser une analogie : imaginons deux ampoules, chacune dans deux maisons différentes. En les intriquant, il devient possible de connaître l'état d'une ampoule (allumée ou éteinte) en observant simplement la seconde, car les deux seraient liées, intriquées. Et cela, immédiatement et même si les maisons sont très éloignées l'une de l'autre.

Ce phénomène d'intrication permet de décrire des corrélations entre les qubits. Si on augmente le nombre de qubits, le nombre de ces corrélations augmente exponentiellement : pour N qubits il y a

$2^n$ corrélations. C'est cette propriété qui confère à l'ordinateur quantique la possibilité d'effectuer des manipulations sur des quantités gigantesques de valeurs, quantités hors d'atteinte d'un ordinateur classique.

Le principe d'incertitude découvert par Werner Heisenberg en 1927, nous dit que, quels que soient nos outils de mesure, nous sommes incapables de déterminer précisément à la fois la position et la vitesse d'un objet quantique (à l'échelle atomique). Soit nous savons précisément où se trouve l'objet et la vitesse semblera fluctuer et devenir floue, soit nous avons une idée précise de la vitesse mais sa position nous échappera.

Julien Bobroff dans son livre La quantique autrement : garanti sans équation ! décrit l'expérience quantique en trois actes :

Le premier moment se situe avant la mesure où l'objet quantique se comporte comme une onde. L'équation de Schrödinger nous permet de prévoir avec précision comment cette dernière se propage, à quelle vitesse, dans quelle direction, si elle s'étale ou se contracte. Ensuite c'est la décohérence qui fait son apparition. La décohérence se passe de manière extrêmement rapide, quasi instantanée. C'est à ce moment précis que l'onde entre en contact avec un outil de mesure (par exemple un écran fluorescent). Cette onde est obligée d'interagir avec les particules qui composent ce dispositif. C'est le moment où l'onde se réduit. La dernière étape, c'est le choix aléatoire parmi tous les états possibles. Le tirage au sort est lié à la forme de la fonction d'onde au moment de la mesure. En fait, seule la forme de la fonction d'onde à la fin du premier acte dicte sa probabilité d'apparaître ici ou là.

Un autre phénomène de la mécanique quantique est l'effet tunnel. Julien Bobroff donne l'exemple d'une balle de tennis. À l'inverse de celle-ci, une fonction d'onde quantique ne rebondit que partiellement contre une barrière. Une petite partie peut pénétrer de l'autre côté par effet tunnel. Cela implique que si la particule est mesurée, elle se matérialisera tantôt à gauche du mur, tantôt à droite.

Un ordinateur quantique utilise donc les lois de la mécanique quantique pour faire des calculs. Il faut qu'il soit dans certaines conditions, parfois extrêmes, comme plonger un système dans de l'hélium liquide pour atteindre des températures proches du zéro absolu soit -273.15°C.

Construire un ordinateur quantique repose sur la capacité de développer une puce informatique sur laquelle sont gravés des qubits. Du point de vue technologique, il existe plusieurs manières de constituer des qubits, nous pouvons utiliser des atomes, des photons, des électrons, des molécules ou des métaux supraconducteurs. Dans certains cas, pour pouvoir fonctionner, un ordinateur quantique a besoin de conditions extrêmes pour opérer comme des températures proches du zéro absolu. Le choix d'IBM par exemple, est d'utiliser des qubits supraconducteurs, construits avec des oxydes d'aluminium (on appelle aussi cette technologie : qubits transmons). Comme évoqué ci-dessus pour permettre et garantir les effets quantiques (superposition et intrication) les qubits doivent être refroidis à une température aussi proche que possible du zéro absolu (soit environ -273°C). Chez IBM ce seuil de fonctionnement est d'environ 20 millikelvins. IBM a démontré la capacité de concevoir un qubit unique en 2007 et en 2016 et a annoncé la mise à disposition dans le cloud d'un premier système physique opérationnel doté de 5 qubits et d'un environnement de développement « QISKit » (Quantum Information Science Kit), permettant de concevoir, tester et optimiser des algorithmes pour des applications commerciales et scientifiques. L'initiative « IBM Q Experience » constitue une première dans le monde industriel.

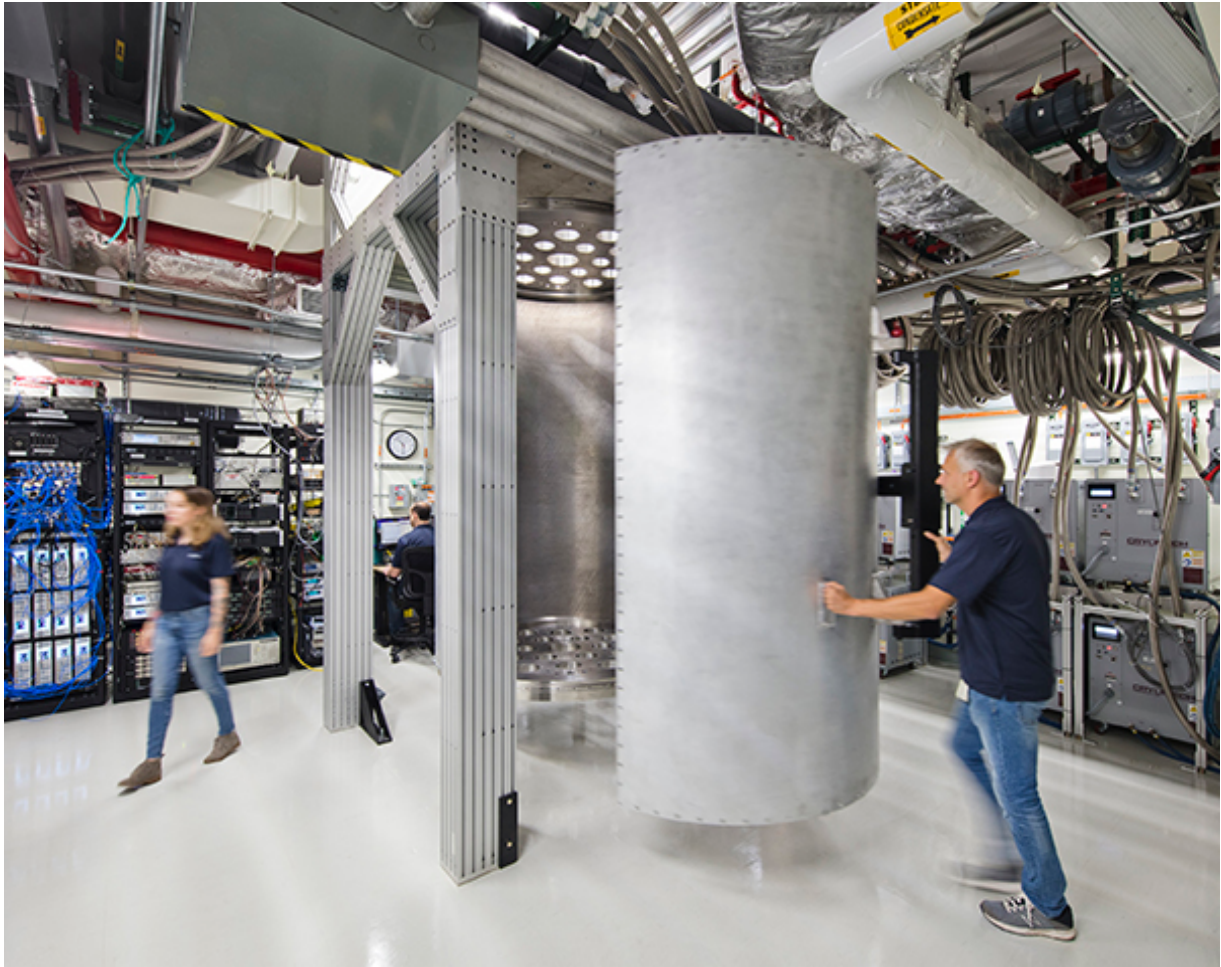
Source : https://www.ibm.com/blogs/research/2020/09/ibm-quantum-roadmap/

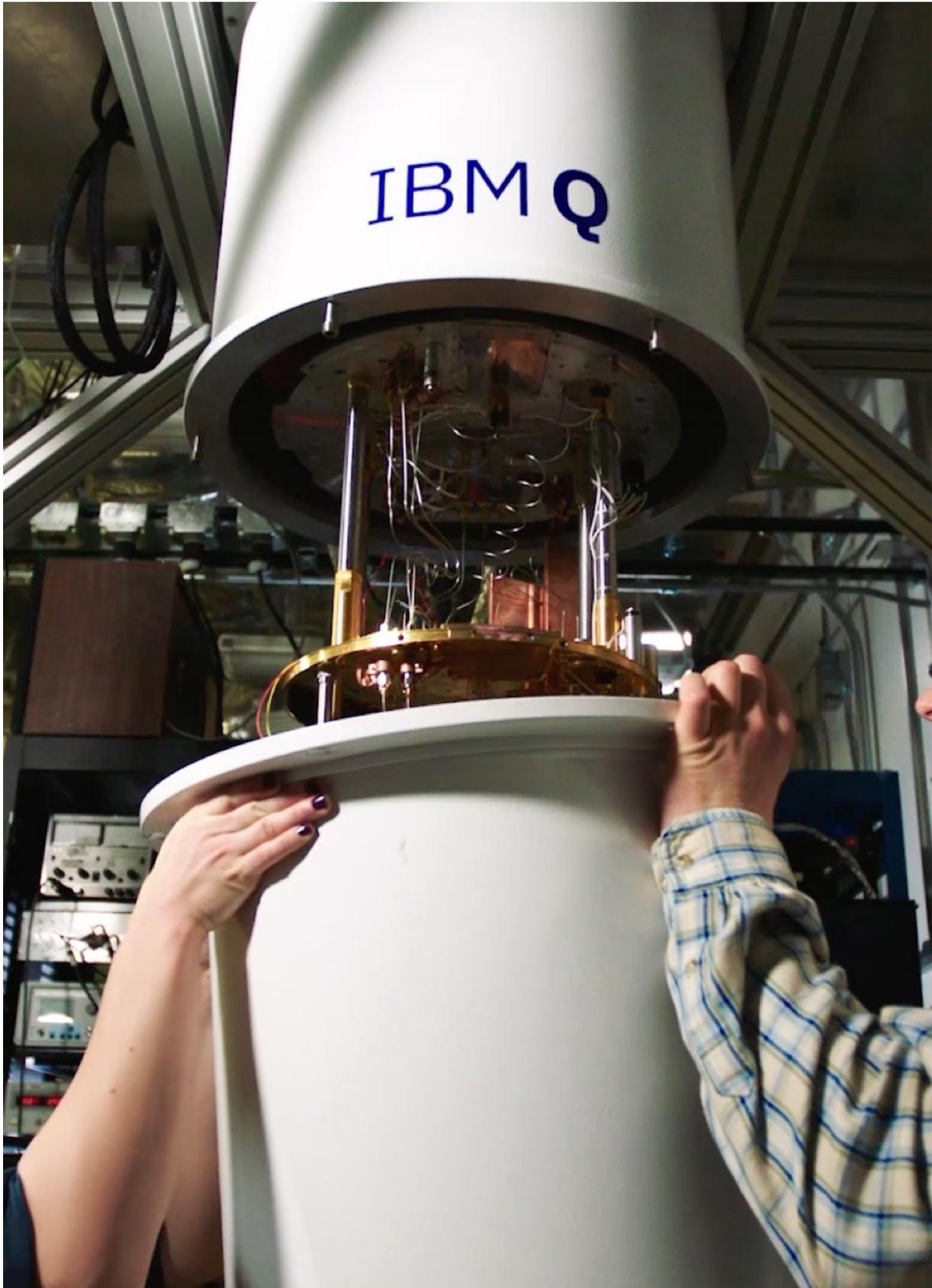

Source : https://www.ibm.com/quantum-computing/ibm-q-network/

IBM a aujourd'hui 32 ordinateurs quantiques, dont un système à 65-qubit, et a récemment publié sa feuille de route.

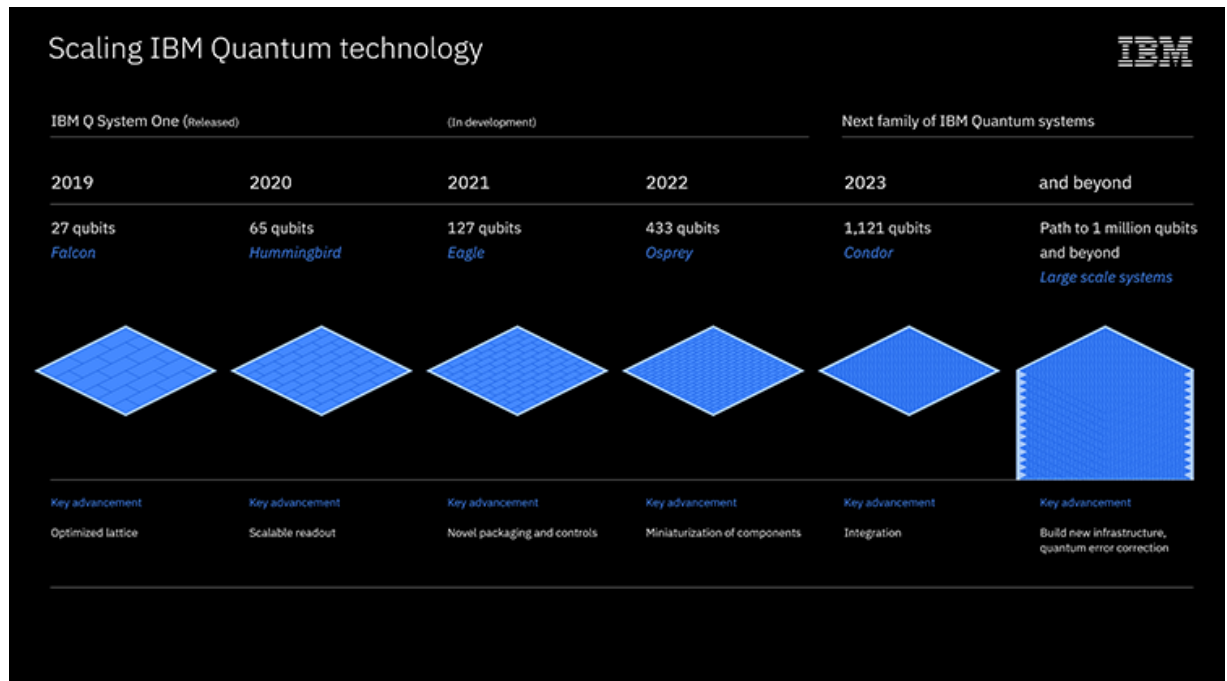
Source : https://www.ibm.com/blogs/research/2020/09/ibm-quantum-roadmap/

Cette ouverture au public a permis à plus de 300 000 utilisateurs de lancer des centaines de milliards d'exécutions de circuits sur du matériel réel et des simulateurs. L'accès à ses machines a conduit à la publication de plus de 400 papiers de recherche par des chercheurs non IBM. IBM a également construit un réseau de plus de 140 membres, que l'on appelle IBM Quantum Network, ayant un accès privilégié aux dernières technologies quantiques pour travailler notamment sur des cas d'usages. IBM Quantum Network est composé de grands groupes, de startups, d'université et laboratoire partout dans le monde.

Le nombre de qubits progressera au fur et à mesure mais cela ne suffit pas. Dans la course au développement d'ordinateurs quantiques, au-delà des qubits, d'autres composantes sont essentielles. On parle de « volume quantique » comme une mesure pertinente de la performance et des progrès technologiques (26). D'autres mesures sont également proposées par des entreprises et laboratoires. On définit aussi « l'avantage quantique » c'est le point à partir duquel les applications du calcul quantique offriront un avantage pratique significatif et démontrable qui dépasse les capacités des seuls ordinateurs classiques. Le concept de volume quantique a été introduit par IBM en 2017. Il commence à se généraliser auprès d'autres constructeurs. Le volume quantique est une mesure qui détermine la puissance d'un système informatique quantique, tenant compte à la fois des erreurs de porte et de mesure, de la diaphonie de l'appareil, ainsi que de la connectivité de l'appareil et de l'efficacité du compilateur de circuits. Le volume quantique est utilisable pour tout système informatique quantique NISQ basé sur des portes et des circuits. Par exemple si vous baissez le taux d'erreurs de x10 sans ajouter de qubits supplémentaires vous pouvez avoir une augmentation du volume quantique de 500x. Au contraire, si vous ajoutez 50 qubits supplémentaires, mais que le taux d'erreur de baisse pas, vous pouvez avoir un volume quantique qui augmente de 0x. L'ajout de qubits ne fait pas tout.

Aujourd'hui, les défis que les chercheurs doivent surmonter sont d'ordre technologique comme la stabilité dans le temps. Quand vous faites tourner un algorithme quantique sur un ordinateur quantique en réel, il y a énormément d'externalités qui peuvent venir perturber l'état quantique de votre programme, qui est déjà fragile. Un autre défi technologique concerne la quantité de qubits que l'on va pouvoir prendre en considération. À chaque fois qu'on augmente la capacité d'un ordinateur quantique d'un qubit, on réduit sa stabilité. Un autre défi, c'est que nous allons être obligés de repenser l'intégralité de l'algorithmie qu'on connaît pour l'adapter à l'informatique quantique. Bien évidemment, il faut pouvoir exécuter des tâches sur ces machines, c'est pourquoi IBM a développé une bibliothèque de programmation spécifique appelée QISKit (Quantum Information Science Kit). Il s'agit d'une librairie open source pour le langage Python, disponible sur qiskit.org. Son développement est très actif, l'ensemble des contributeurs, dont IBM, fait régulièrement évoluer les fonctionnalités de cet environnement de programmation.

Les ordinateurs quantiques vont s'ajouter aux ordinateurs classiques pour adresser des problèmes qui sont aujourd'hui non résolus. Les ordinateurs classiques peuvent par exemple calculer des problèmes complexes que ne peut calculer un ordinateur quantique. Il y a des problèmes que les deux ordinateurs, classique et quantique, pourront résoudre. Et enfin, des défis que ne peut résoudre un ordinateur classique mais qu'un ordinateur quantique pourra adresser. Beaucoup d'applications sont possibles dans le domaine de la chimie, la science des matériaux, le machine learning ou l'optimisation.

Par exemple, il est difficile pour un ordinateur classique de calculer de façon exacte (c'est-à-dire sans aucune approximation) l'énergie de la molécule de caféine, pourtant de taille moyenne avec 24 d'atomes, c'est un problème très complexe (27) (28). Nous aurions approximativement besoin de $10^{48}$ bits pour représenter la configuration énergétique d'une seule molécule de caféine à un instant t. Soit presque le nombre d'atomes admis sur terre qui est de $10^{50}$. Mais nous pensons qu'il est possible de le faire avec 160 qubits.

Mais ce n'est pas la seule limite. Pour vous donner une illustration simple, nous pouvons parler du problème dit « du vendeur itinérant », ou plus exactement aujourd'hui, le problème de l'acheminement des camions de livraison. Si vous souhaitez choisir l'itinéraire le plus efficace pour qu'un camion livre des colis à cinq adresses, il existe 12 itinéraires possibles, il est donc au moins possible d'identifier le meilleur. Cependant, à mesure que vous ajoutez d'autres adresses, le problème devient exponentiellement plus difficile - au moment où vous avez 15 livraisons, il y a plus de 43 milliards d'itinéraires possibles, ce qui rend pratiquement impossible de trouver le meilleur. Par exemple, pour 71 villes, le nombre de chemins candidats est supérieur à $5 \times 10^{80}$.

À l'heure actuelle, l'informatique quantique est adaptée pour certains algorithmes comme l'optimisation, le Machine Learning ou la simulation. Avec ce type d'algorithmes, les cas d'usage s'appliquent dans plusieurs secteurs industriels. Les services financiers comme l'optimisation des risques du portefeuille, la détection des fraudes, la santé (recherche de médicaments, étude des protéines, etc.), les chaînes d'approvisionnement et la logistique, la chimie, la recherche de nouveaux matériaux ou le pétrole sont autant de domaines qui vont être prioritairement impactés.

Nous pouvons aborder plus spécifiquement le futur de la recherche médicale avec le quantique qui devrait permettre à terme de synthétiser de nouvelles molécules thérapeutiques. Si nous voulons relever le défi du changement climatique par exemple, nous devons résoudre de nombreux problèmes tels que concevoir de meilleures batteries, trouver des moyens moins énergivores de cultiver nos aliments et planifier nos chaînes d'approvisionnement pour minimiser les transports. Résoudre efficacement ces problèmes nécessite des approches informatiques radicalement améliorées dans des domaines tels que la chimie et la science des matériaux, ainsi que dans l'optimisation et la simulation - des domaines où l'informatique classique est confrontée à de sérieuses limitations. Pour un ordinateur classique faire le produit de deux nombres et obtenir le résultat est une opération très simple : 7*3 = 21 ou 6739*892721 = 6016046819. Cela reste vrai pour de très grands nombres. Mais le problème inverse est nettement plus complexe. Connaissant un grand nombre N, il est plus compliqué de trouver P et Q tel que : P x Q = N. C'est cette difficulté qui est à la base des techniques de cryptographie courantes. Pour un tel problème, à titre d'exemple on estime qu'un problème qui durerait 1025 jours sur un ordinateur classique pourrait être résolu en quelques dizaines de secondes sur une machine quantique. On parle pour ce cas d'accélération exponentielle. Avec les ordinateurs quantiques, nous pouvons aborder les problèmes de manière entièrement nouvelle en tirant parti de l'intrication, de la superposition et des interférences : modéliser les procédures physiques de la nature, effectuer beaucoup plus simulations de scénarios, obtenir de meilleures solutions d'optimisation, trouver de meilleurs modèles dans les processus AI / ML. Dans ces catégories de problème éligibles aux ordinateurs quantiques, on trouve beaucoup de cas d'optimisation, dans les domaines logistiques (plus court chemin), de la finance (estimation de risques, évaluation de portefeuilles d'actifs), du marketing (« maxcut », « clique »), de l'industrie et de la conception de systèmes complexes (satisfiabilité, parcours de graphes) (29) (30) (31) (32) (33) (34). Le domaine de l'IA (35) (36) est également un champ de recherche actif, et des méthodes d'apprentissage pour les réseaux de neurones artificiels commencent à voir le jour, c'est donc l'ensemble des activités humaines concernées par le traitement de l'information qui sont potentiellement concernées par l'avenir du calcul quantique. Le domaine de la cybersécurité et de la cryptographie est également un sujet d'attention. L'algorithme de Shor a été démontré voici plus de 20 ans et il pourrait rendre fragile le chiffrage communément utilisé sur internet. Il faudra attendre que les machines quantiques soient suffisamment puissantes pour traiter ce type de calcul particulier, et d'autre part des solutions de cryptage sont déjà connues et démontrées hors d'atteinte de cet algorithme. Parmi les solutions, il y a les technologies quantiques elles-mêmes qui permettent de générer et de transporter des clefs de cryptage de manière absolument inviolable. De ce fait le domaine des technologies quantiques et du calcul quantique en particulier est considéré comme un enjeu stratégique, et l'Europe, la France et bien d'autres pays soutiennent les efforts de recherche dans ce domaine.

Si l'on prend les cas d'usages par secteurs, nous pouvons en trouver de nombreux dans les banques et institutions financières : améliorer les stratégies de trading, améliorer les portefeuilles clients et mieux analyser les risques financiers. Un algorithme quantique en cours de développement, par exemple, pourrait potentiellement fournir une accélération quadratique dans les cas d'utilisation de la tarification des produits dérivés - un instrument financier complexe qui nécessite 10 000 simulations pour être évalué sur un ordinateur classique ne nécessiterait que 100 opérations quantiques sur un appareil quantique.

Un des cas d'usages du quantique est l'optimisation du trading. Il sera également possible pour les banques d'accélérer les optimisations de portefeuille comme les simulations de Monte-Carlo. La simulation d'achat et vente de produits (trading) tels que les produits dérivés peuvent être améliorés grâce à l'informatique quantique. La complexité des activités de trading sur les marchés financiers monte en flèche. Les gestionnaires de placements peinent à intégrer des contraintes réelles, telles que la volatilité du marché et les changements des événements de la vie des clients, dans l'optimisation du portefeuille. Actuellement, le rééquilibrage des portefeuilles d'investissement qui suivent les mouvements du marché est fortement impacté par les contraintes de calcul et les coûts de transaction. La technologie quantique pourrait aider à réduire la complexité des environnements commerciaux d'aujourd'hui. Les capacités d'optimisation combinatoire de l'informatique quantique peuvent permettre aux gestionnaires de placements d'améliorer la diversification du portefeuille, de rééquilibrer les investissements de portefeuille pour répondre plus précisément aux conditions du marché et aux objectifs des investisseurs, et de rationaliser de manière plus rentable les processus de règlement des transactions. Le machine learning est aussi utilisé pour l'optimisation de portefeuille et la simulation de scénarii. Les banques et les institutions financières comme les hedge funds sont de plus en plus intéressées, car ils y voient une façon de minimiser les risques tout en maximisant les gains avec des produits dynamiques qui s'adaptent en fonction de nouvelles données simulées. La finance personnalisée est aussi un domaine exploré. Les clients exigent des produits et services personnalisés qui anticipent rapidement l'évolution de leurs besoins et de leurs comportements. De petites et moyennes institutions financières peuvent perdre des clients en raison d'offres qui ne privilégient pas l'expérience client. Il est difficile de créer des modèles analytiques qui passent au crible des monticules de données comportementales suffisamment rapidement et avec précision pour cibler et prévoir les produits dont certains clients ont besoin en temps quasi réel. Un problème similaire existe dans la détection des fraudes pour trouver des modèles de comportements inhabituels. On estime que les institutions financières perdent entre 10 et 40 milliards de dollars de revenus par an en raison de fraudes et de mauvaises pratiques de gestion des données. Pour le ciblage client et la modélisation des prévisions, l'informatique quantique pourrait changer la donne. Les capacités de modélisation des données des ordinateurs quantiques devraient s'avérer supérieures pour trouver des modèles, effectuer des classifications et faire des prédictions qui ne sont pas possibles aujourd'hui avec les ordinateurs classiques en raison des défis des structures de données complexes. Un autre cas d'usage dans le monde de la finance est l'analyse de risques. Les calculs de l'analyse des risques sont difficiles, car il est difficile sur le plan informatique d'analyser de nombreux scénarios. Les coûts de mise en conformité devraient plus que doubler au cours des prochaines années. Les institutions de services financiers sont soumises à des pressions croissantes pour équilibrer les risques, couvrir les positions plus efficacement et effectuer un plus large éventail de tests de résistance pour se conformer aux exigences réglementaires. Aujourd'hui, les simulations de Monte-Carlo - la technique privilégiée pour analyser l'impact du risque et de l'incertitude dans les modèles financiers - sont limitées par la mise à l'échelle de l'erreur d'estimation. Les ordinateurs quantiques ont le potentiel d'échantillonner les données différemment en testant plus de résultats avec une plus grande précision, fournissant une accélération quadratique pour ces types de simulations.

La modélisation moléculaire permet des découvertes telles que des batteries au lithium plus efficaces. Pouvoir modéliser les processus physiques de la nature permettra de développer de nouveaux matériaux avec des propriétés différentes. L'informatique quantique donnera les moyens de simuler les atomes ou interactions atomiques beaucoup plus précisément et à des échelles beaucoup plus grandes. Des nouveaux matériaux vont pouvoir être utilisés partout que cela soit dans les produits de consommation ou les voitures, les batteries, etc. L'informatique quantique permettra d'exécuter des calculs d'orbite moléculaire sans approximations. Une autre application est l'optimisation du réseau électrique d'un pays, une modélisation environnementale plus prédictive et la recherche de sources d'énergie à émissions plus faibles. L'aéronautique sera également source de cas d'usages. Dans le transport aérien, à chaque atterrissage d'un avion des centaines d'opérations se mettent en place : Changement d'équipage, pleins de carburant, nettoyage de la cabine, livraison des bagages, inspections des éléments moteurs et train d'atterrissage. Chaque opération fait elle-même appel à des sous-opérations (pour le plein de carburant, il faut un camion-citerne disponible, un chauffeur du camion et 2 personnes pour le remplissage. Au préalable il faut bien sûr remplir la citerne, etc.). Donc au total des centaines d'opérations/actions élémentaires à prévoir, et cela pour tous les avions atterrissant dans le même créneau horaire. Imaginons un orage sur Paris retarde de 30mn l'atterrissage. Il faut alors en temps réel tout recalculer pour tous les avions. Les véhicules électriques ont une faiblesse : la capacité et la vitesse de charge de leurs batteries. Une percée en informatique quantique réalisée par des chercheurs d'IBM et de Daimler AG (37), la société mère de Mercedes-Benz, pourrait aider à relever ce défi. Le constructeur automobile Daimler, s'intéresse de près à l'impact de l'informatique quantique sur l'optimisation de la logistique des transports, jusqu'aux prédictions sur les futurs matériaux pour la mobilité électrique, notamment la prochaine génération de batteries. Il y a tout lieu d'espérer que les ordinateurs quantiques donneront des résultats initiaux dans les années à venir pour simuler avec précision la chimie des cellules de batterie, les processus de vieillissement et les limites de performance des cellules de batterie. Lorsque les temps d'exécution d'une solution précise prennent trop de temps, les entreprises se contentent de calculs inférieurs. Le problème du voyageur du commerce peut être étendu dans de nombreux domaines comme l'énergie, les télécommunications, la logistique, les chaines de productions ou l'allocation de ressources. Par exemple, dans le fret maritime, il y a une grande complexité dans la gestion des containers de bout en bout : charger, convoyer, livrer puis décharger dans plusieurs ports dans le monde est un problème multiparamétrique pouvant être adressé par l'informatique quantique. Une meilleure compréhension des interactions entre les atomes et les molécules permettra de découvrir de nouveaux médicaments. L'analyse en détail des séquences d'ADN permettra de détecter les cancers plus tôt en développant des modèles qui détermineront comment les maladies se développent. L'avantage du quantique sera d'analyser dans le détail à une échelle jamais atteinte le comportement des molécules. Les simulations de scénarios permettront de mieux prédire les risques d'une maladie ou sa diffusion, la résolution de problèmes d'optimisations permettra d'optimiser les chaînes de distributions des médicaments, et enfin l'utilisation de l'IA permettra d'accélérer les diagnostics, d'analyser plus précisément les données génétiques.

## Conclusion

Le centre de données de demain est un centre de données composé de systèmes hétérogènes, qui tourneront des charges de travail hétérogènes. Les systèmes seront localisés au plus près des données. Les systèmes hétérogènes seront dotés d'accélérateurs binaires, d'inspirations biologiques et quantiques.  Tel un chef d'orchestre, le cloud hybride permettra grâce à une couche de sécurité et d'automatisation intelligente de mettre ces systèmes en musique.